**FRONT MATTER**

**Title**

- Addressable metasurfaces for dynamic holography and optical information encryption

**Authors**

Jianxiong Li,[1] Simon Kamin,[1] Guoxing Zheng,[2] Frank Neubrech,[3] Shuang Zhang,[4*] and Laura Na Liu[1,3*]

**Affiliations**

[1]Max Planck Institute for Intelligent Systems, Heisenbergstrasse 3, 70569 Stuttgart, Germany.

[2]School of Electronic Information, Wuhan University, Wuhan 430072, China.

[3]Kirchhoff Institute for Physics, University of Heidelberg, Im Neuenheimer Feld 227, 69120 Heidelberg, Germany.

[4]School of Physics & Astronomy, University of Birmingham, Birmingham B15 2TT, UK.

[*]Corresponding author. E-mail: laura.liu@is.mpg.de, s.zhang@bham.ac.uk.

**Abstract**

Metasurfaces enable manipulation of light propagation at an unprecedented level, benefitting from a number of merits unavailable to conventional optical elements, such as ultracompactness, precise phase and polarization control at deep subwavelength scale, and multi-functionalities. Recent progress in this field has witnessed a plethora of functional metasurfaces, ranging from lenses and vortex beam generation to holography. However, research endeavors have been mainly devoted to static devices, exploiting only a glimpse of opportunities that metasurfaces can offer. Here, we demonstrate a dynamic metasurface platform, which allows for independent manipulation of addressable subwavelength pixels at visible frequencies through controlled chemical reactions. In particular, we create dynamic metasurface holograms for advanced optical information processing and encryption. Plasmonic nanorods tailored to exhibit hierarchical reaction kinetics upon hydrogenation/dehydrogenation constitute addressable pixels in multiplexed metasufaces. The helicity of light, hydrogen, oxygen, and reaction duration serve as multiple keys to encrypt the metasurfaces. Remarkably, one single metasurface can be deciphered into manifold messages with customized keys, featuring a compact data storage scheme as well as a high level of information security. Our work suggests a novel route to protect and transmit classified data, where highly restricted access of information is imposed.

Teaser: We present addressable optical metasurfaces for optical information encryption and holography

**MAIN TEXT**

**Introduction**

Metasurfaces outline a new generation of ultrathin optical elements with exceptional control over the propagation of light(*1-11*). In particular, metasurface holograms have attracted tremendous attention because they promise useful applications in displays, security, and data storage(*12, 13*). While conventional phase holography acquires phase modulation via propagation through wavelength-scale unit cells with different heights or different refractive indices, metasurface holography relies on the spatial variation of phase discontinuities induced by a single-layer array composed of subwavelength antennas(*1, 3*). Especially, geometric metasurface holography that shapes light wavefronts via the Pancharatnam-Berry (PB) phase can be achieved by simply controlling the in-plane orientations of antennas. This approach not only allows for highly precise control of the phase profile, but also alleviates the fabrication complexity. More importantly, the PB phase does not depend on the specific antenna design or wavelength, rendering a broadband performance possible(*12, 14*).

To date, a number of metasurface holograms have been accomplished for the terahertz, infrared, and visible spectral regimes(*15-17*). However, most of the reconstructed holographic images have been restricted to be static due to the fixed phase and amplitude profiles possessed by the metasurfaces once fabricated(*14, 18, 19*). Only very recently, attempts on stretchable metasurface holograms have been made, showing image switching at different planes(*20*). In addition, electrically tunable metasurface holograms have been demonstrated at microwave frequencies(*21*). Such initial endeavors elicit an inevitable transition from static to dynamic metasurface holography for advancing the field forward(*22*). Nevertheless, significant challenges remain for imparting complex reconfigurablity at visible frequencies. The challenges include multiplexing a metasurface with addressable subwavelength pixels, endowing the pixels with different dynamic functions, independent control of the dynamic pixels at the same time, and so forth. In this Letter, we overcome these challenges by employing chemically active metasurfaces based on magnesium (Mg), which comprise dynamic plasmonic pixels enabled by the unique hydrogenation/dehydrogenation characteristics of Mg. In particular, we demonstrate a series of dynamic metasurface holograms at visible frequencies with novel functionalities including a hologram with both static and dynamic patterns, a hologram with differently sequenced dynamics, and an encrypted hologram that can be deciphered into manifold customized messages.

**Results**

**Phase-transition of Mg nanorod**

Mg not only possesses excellent plasmonic properties at visible frequencies, but also can undergo a phase-transition from metal to dielectric upon hydrogen loading, forming magnesium hydride ($MgH_2$) (see Fig. 1A)(*23-25*). This transition is reversible through dehydrogenation using oxygen. As a result, the plasmonic response of an Mg nanorod can be reversibly switched on and off, constituting a dynamic plasmonic pixel as shown in Fig. 1B (see also Fig. S1). To facilitate hydrogenation and dehydrogenation processes, a titanium (Ti, 5 nm) spacer and a palladium (Pd, 10 nm) catalytic layer are capped on the Mg nanorods, which reside on a silicon dioxide film (100 nm) supported by a silicon substrate. A Ti (3 nm) adhesion layer is adopted underneath for improved Mg quality. For conciseness, Ti is omitted in all the figure illustrations. Such plasmonic pixels can be employed to dynamically control the phase of circularly polarized light via the PB phase. To shape arbitrary light wavefronts, we choose 8-phase levels for the metasurfaces in this work. The $n^{th}$ phase pixel $\varphi_n$ is represented by a nanorod with an orientation angle $\varphi_n/2$ as shown in Fig. 1C.

**Dynamic metasurface holograms**

Metasurfaces multiplexed with addressable plasmonic pixels render encoding of manifold optical information with tailored dynamic possibilities. To show such unprecedented degrees of freedom, we demonstrate a series of dynamic metasurface holograms for optical information processing and encryption at visible frequencies. As the first example, the metasurface is multiplexed with two independent phase profiles containing Mg/Pd ($P_1$) and Au ($P_2$) nanorods as dynamic and static pixels, respectively. Two off-axis holographic images with Chinese letters, 'harmony' (left) and 'peace' (right) are respectively encoded into each phase profile based on Gerchberg-Saxton algorithm(*26*) as shown in Fig. 2A. The two sets of pixels are interpolated into each other with a displacement vector of (300 nm, 300 nm) as shown by the scanning electron microscopy (SEM) image in Fig. 2B. Fig. 2C illustrates the time evolutions of the scattered intensities of $P_1$ and $P_2$ recorded at their resonance positions during hydrogenation (10% hydrogen) and dehydrogenation (20% oxygen), respectively (see also Fig. S1). Upon hydrogen exposure at $t_1$, the scattered intensity of $P_1$ instantly decreases and reaches the 'off' state at $t_2$ as depicted by the solid line. Upon oxygen exposure, $P_1$ can be switched on again at $t_3$, exhibiting slight hysteresis in the recovered scattered intensity (see Fig. 2C and Fig. S2). In contrast, the scattered intensity of $P_2$ keeps constant as depicted by the dashed line in Fig. 2C, remaining at the 'on' state during the entire process. Such distinct optical responses of $P_1$ and $P_2$ enable dynamic and static holographic patterns that can be reconstructed from the same metasurface.

The experimental set-up for the hologram characterizations is presented in Fig. 2D, in which circular polarized light is generated from a laser diode source (633 nm) by a polarizer and a quarter-wave plate. The light is incident onto the metasurface sample placed in a gas cell. The reflected hologram is projected onto a screen in the far field and captured by a visible camera. Fig. 2E shows the representative snapshots of the holographic images during hydrogenation and dehydrogenation. At the initial state, two high-quality holographic patterns 'harmony' (left) and 'peace' (right) are observed by illumination of right circularly polarized (RCP) light onto the metasurface sample (see Fig. S3A for the enlarged holographic pattern). When left circularly polarized (LCP) light is applied, the positions of the two images are swapped (not shown). Upon hydrogenation, 'harmony' gradually diminishes, whereas 'peace' remains unchanged. Upon dehydrogenation, 'harmony' can be recovered so that both holographic patterns are at the 'on' state again. A video that records the dynamic evolution of the metasurface hologram can be found in Movie S1a. To demonstrate good reversibility and durability, operation of the metasurface hologram in a number of cycles is shown in Movie S1b.

To achieve holographic patterns with sequenced dynamics, the metasurface is multiplexed with dynamic pixels that possess different reaction kinetics upon hydrogenation/dehydrogenation (see Fig. S4A). As shown by the SEM image in Fig. 3A, each unit cell contains a Mg/Pd ($P_1$) nanorod and a Mg/Pd/Cr ($P_3$) nanorod as two sets of dynamic pixels. The Cr (1 nm) capping layer can effectively slow down both the hydrogenation and dehydrogenation rates of $P_3$. This leads to distinct time evolutions of $P_1$ (solid line) and $P_3$ (dash-dotted line) during hydrogenation and dehydrogenation (see Fig. 3B). Specifically, upon hydrogen exposure at $t_1$ the scattered intensity of $P_1$ decreases more rapidly than that of $P_3$, quickly reaching the 'off' state at $t_2$, whereas $P_3$ is only switched off after a much longer time $t_3$, exhibiting an evident delay. During dehydrogenation, $P_1$ can be promptly switched on at $t_4$, whereas $P_3$ is gradually recovered to the 'on' state until $t_5$. As a result, $P_1$ and $P_3$ can be switched on and off independently. The two holographic patterns reconstructed using RCP light, *i.e.*, the portrait of Marie Curie as well as the chemical elements Po and Ra, can transit through four distinct states as shown in Fig. 3C (see Fig. S3B for the enlarged holographic pattern). $H_2$ (1) or $H_2$ (2) refers to a hydrogenation process until the first or the second feature completes transition. A video that records the dynamic evolution of the metasurface hologram can be found in Movie S2.

**Optical information processing and encryption**

The ability to endow addressable plasmonic pixels with hierarchical dynamics in a programmable manner allows for advanced optical information transmission and encryption. As a proof-of-concept experiment, a set of geometric codes have been designed and illustrated in Fig. 4A. Each Arabic number is represented by a symbol, comprising a unique sequence of dots and dashes, similar to the coding principle of Morse codes. All the numbers from 0 to 9 can be expressed by specific holographic images that are reconstructed from the same metasurface based on the mechanism as shown in Fig. 4B. Each unit cell contains three different pixels, $P_1$ (Mg/Pd), $P_2$ (Au), and $P_3$ (Mg/Pd/Cr). The holographic patterns achieved by $P_1$ and $P_3$ are hollow dashes at two different locations, whereas the holographic pattern generated by $P_2$ comprises one solid dash and three dots (see also Fig. S4B). The shape of the dots is complementary to that of the hollow dashes so that they can fit exactly in space to form two solid dashes. It is noteworthy that the intensities of the individual holographic patterns are inversely proportional to their pattern areas. Therefore, $P_2$ is applied twice in each unit cell for achieving a uniform intensity distribution within a merged hologram. The reconstructed holographic image of '1' is presented in Fig. 4B. The holographic images of other numbers can be generated through different routes governed by the helicity of light, the sequences of hydrogenation and dehydrogenation, as well as the reaction duration (see Fig. 4C). We define a coding rule that each number corresponds to a union set of the two holographic patterns in zones I and II as shown in Fig. 4C. It is known that the sign of the acquired phase profile is flipped, when the helicity of the incident light is changed(*12*). Therefore, two centrosymmetric holographic patterns can occur in zones I and II, respectively. When $P_1$, $P_2$ and $P_3$ are all at the 'on' state, RCP light illumination uncovers '1' in zone II, whereas LCP light illumination gives rise to its centrosymmetric image in zone I, corresponding to '9' (see Fig. 4C as well as Fig. S5, A and B). The centrosymmetry point is indicated by a white circle in each plot, highlighting the location of the zero-order reflected light. When linearly polarized (LP) light is applied, the two holographic patterns appear simultaneously. Based on the coding rule, a union set of the holographic patterns in the two zones gives rise to four solid dashes, corresponding to '0' (see Fig. 4C and Fig. S5C). Through sequenced hydrogenation and dehydrogenation, the geometric codes corresponding to the rest of the numbers can be achieved as shown in Fig. 4C. A video that records the dynamic transformation among different numbers can be found in Movies S3a and S3b.

Following the Morse code abbreviations, in which different combinations of numbers represent diverse text phrases for data transmission, we demonstrate a highly secure scheme for optical information encryption and decryption using dynamic metasurfaces. As illustrated in Fig. 5, Alice would like to send different messages to multiple receivers including Tim, Bob, Ted, and so forth. These messages are all encrypted on one metasurface. Identical samples are sent to the receivers. Upon receipt of his sample together with the customized keys, Tim reads out the information of 88, which means 'love and kisses' in Morse codes following LCP/$H_2$(1). Bob first applies LCP/$H_2$(1)/$H_2$(2)/$O_2$ to obtain '7'. After resetting the sample through sufficient dehydrogenation, he reads out '3' following RCP/$H_2$(1)/$H_2$(2)/$O_2$. The decrypted message is therefore 'best regards'. Similarly, other receivers can decode their respective messages using the provided keys. In other words, a single metasurface design can be deciphered into manifold holographic messages, given the sequences of keys are customized. This elucidates an unprecedented level of data compactness and security for transmitting information especially to a pool of multiple receivers.

**Discussion**

If not only the keys but also the metasurfaces are customized, the conveyed data quantity can be dramatically increased. Also, dynamic pixels capped with Cr of different thicknesses in a unit cell may introduce holographic patterns with more dynamic hierarchies. Furthermore, a variety of holographic patterns including numbers, letters, pictures, and among others can be combined to encrypt a significant amount of complex information with a higher security level. Therefore, our technique, which enables independent control of addressable dynamic pixels, can lead to novel data storage and optical communication systems with high spatial resolution and high data density using ultra-thin devices. This will be very useful for modern cryptography and security applications. In addition, our scheme can be readily applied to achieve compact optical elements for dynamic beam steering, focusing and shaping and dynamic optical vortex generations, largely enriching the functionality breadth of current metasurface systems.

**Materials and Methods**

**Structure fabrication**

The samples in Fig. 2 were fabricated using multi-step electron-beam lithography (EBL). First, a structural layer composed of Au nanorods and alignment markers were defined in a PMMA resist (Allresist) using EBL (Raith e_line) on a $SiO_2$ (100 nm)/Si substrate. A 2 nm chromium adhesion layer and a 50 nm Au film were successively deposited on the substrate through thermal evaporation followed by metal lift-off. The dimension of the Au nanorod is 200 nm × 80 nm × 50 nm. Next, the substrate was coated with a PMMA resist. Computer-controlled alignment using the Au markers was carried out to define a second structural layer. Subsequently, 3 nm Ti, 50 nm Mg, 5 nm Ti, and 10 nm Pd multilayers were deposited on the substrate through electron-beam evaporation followed by metal lift-off. The samples in Figs. 3 and 4 were manufactured following similar procedures but with different metal depositions as described in the main text.

**Optical measurements**

The scattering spectra of a single Mg nanorod before and after hydrogenation were measured using a microspectrometer (Princeton Instruments, Acton SP-2356 Spectrograph with Pixis:256E silicon CCD camera) through a NT&C dark-field microscopy set-up (Nikon ECLIPSE LV100ND microscope and Energetiq Laser-Driven Light Source, EQ-99). A polarizer was used to generate linearly polarized light with polarization parallel to the long axis of the Mg nanorod for the scattering spectrum measurements. The scattering spectra were normalized with respect to that of a bare substrate. The hydrogenation and dehydrogenation experiments were carried out in a homebuilt stainless-steel chamber.

Ultrahigh-purity hydrogen (UHP 5.0), oxygen (UHP 5.0) and nitrogen (UHP 5.0) from Westfalen were used with FSP-1246A mass-flow controllers (FLUSYS GmbH, Offenbach, Germany) to adjust the flow rates and gas concentrations in the chamber. In this work, the hydrogenation ($P_1$ and $P_3$) and dehydrogenation ($P_1$) processes were carried out at room temperature. Dehydrogenation of $P_3$ was carried out at 80 °C for facilitating the process. The flow rate of hydrogen and oxygen was 2.0 l/min.

**Numerical simulations**

Numerical simulations of the phase delay of the Mg nanorods were carried out using commercial software COMSOL Multiphysics based on a finite element method. Periodic boundary conditions, waveguide port boundary conditions, and perfectly matched layers were used for calculations of the structure arrays. The substrate was included in the simulations. The refractive index of $SiO_2$ was taken as 1.5. The dielectric constants of Si and Mg were taken from Palik(*27*). The dielectric constants of Pd and Ti were taken from Rottkay(*28*).

**Design of the metasurface holograms**

To generate a target image, a phase-only hologram with a unit cell size of 600 nm and a periodicity of 600 μm along both directions, respectively, was designed based on Gerchberg-Saxton algorithm. The off-axis angle of the reconstructed hologram was 17°. A 2×2 periodic array of the holographic pattern was used to avoid formation of laser speckles in the holographic image, according to the concept of Dammann gratings(*29*). Owing to the large angular range, the Rayleigh–Sommerfeld diffraction method was used to simulate the holographic image(*30*). The hologram was pre-compensated to avoid pattern distortions.

**H2: Supplementary Materials**

Fig. S1. Dynamic scattering spectra of an Mg nanorod during hydrogenation and dehydrogenation.

Fig. S2. High quality and high resolution of the holographic patterns.

Fig. S3. Principle of the multiplexed metasurfaces.

Fig. S4. Reconstructed holographic patterns in different zones upon LCP, RCP, or LP light.

Movie S1. Evolution of the metasurface hologram with static and dynamic states.

Movie S2. Evolution of the metasurface hologram with sequenced dynamics.

Movie S3. Dynamic transformation among different numbers.

**Acknowledgements**

**General:** We thank the 4th Physics Institute at the University of Stuttgart for kind permission to use their electron-gun evaporation system. We thank N. Gladen and M. Matuschek for help with material processing as well as X. Y. Duan for video programming. **Funding:** This project was supported by the Sofja Kovalevskaja grant from the Alexander von Humboldt-Foundation and the European Research Council (ERC Dynamic Nano and ERC Topological) grants. We gratefully acknowledge the generous support by the Max-Planck Institute for Solid State Research for the usage of clean room facilities. **Author contributions:** J.X.L. and N.L. conceived the project. S. Z. provided crucial suggestions to the main concept of the project. J.X.L. performed the experiments and theoretical calculations. S.K. built the gas system, and helped with the Mg material deposition. G.X.Z. helped with the hologram designs. F.N. made helpful comments to the manuscript. **Competing interests:** The authors declare no competing financial interest. **Data and materials availability:** The data that support the plots within this paper and other findings of this study are available from the corresponding author upon reasonable request.

**Figures**

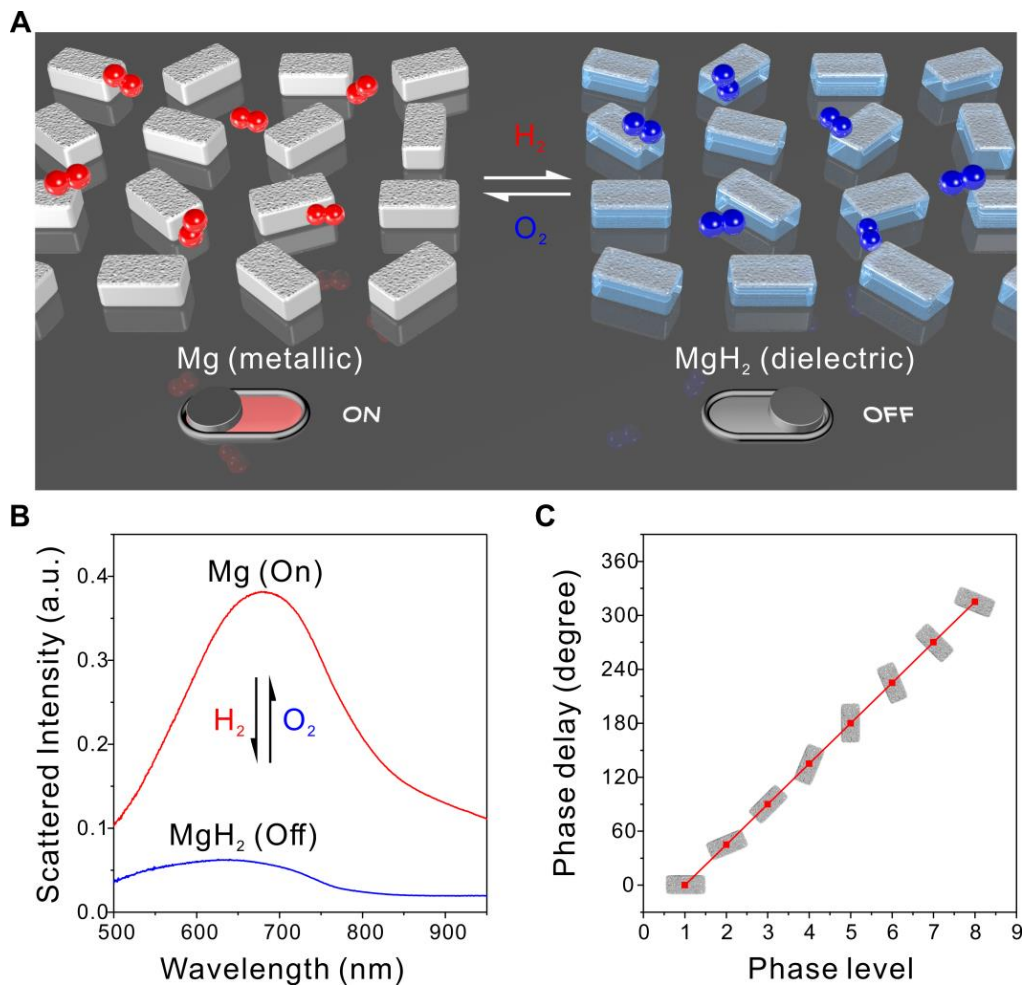

**Fig. 1. Working principle of the dynamic metasurface pixels**. (**A**) Schematic of hydrogen-responsive Mg nanorods (200 nm × 80 nm × 50 nm), which are sandwiched between a Ti (5 nm)/Pd (10 nm) capping layer and a Ti (3 nm) adhesion layer. (**B**) Measured scattering spectra of such a nanorod in Mg (on) and $MgH_2$ (off) states. Before hydrogenation, the Mg nanorod exhibits a strong plasmonic resonance (red curve), whereas after hydrogenation (10% $H_2$) the $MgH_2$ rod shows a nearly featureless spectrum (blue curve). The process is reversible through dehydrogenation using $O_2$ (20%). (**C**) Simulated phase delay for the different phase levels. The orientation of the nanorod at each phase level is shown.

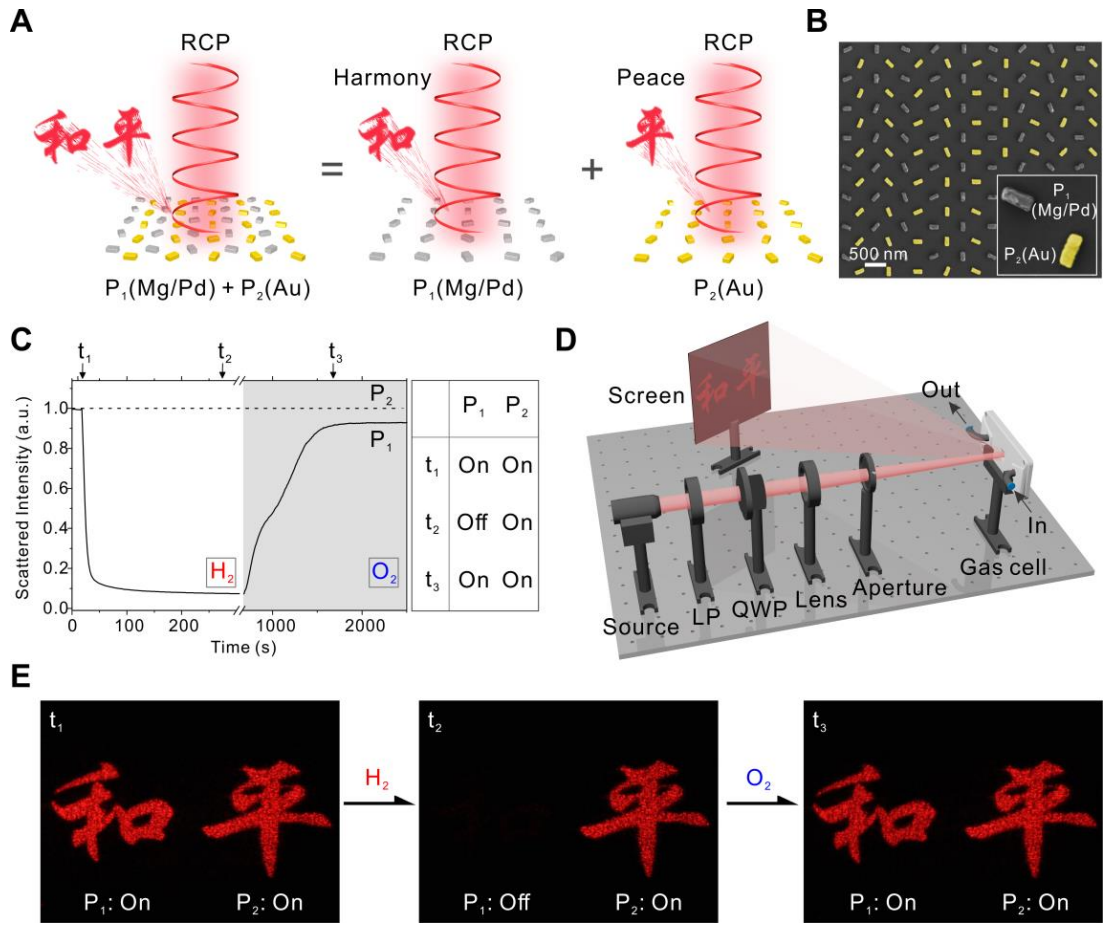

**Fig. 2. Metasurface hologram with dynamic and static patterns. (A)** Two holographic patterns 'harmony' and 'peace' reconstructed from two independent phase profiles containing Mg/Pd ($P_1$) and Au ($P_2$) nanorods as dynamic and static pixels, respectively. Ti is omitted for conciseness. The two sets of pixels are merged together with a displacement vector of (300 nm, 300 nm). **(B)** Overview SEM image of the hybrid plasmonic metasurface. The Au ($P_2$) nanorods are represented in yellow color. Inset: enlarged SEM image of the unit cell (600 nm × 600 nm). **(C)** Time evolutions of $P_1$ (solid line) and $P_2$ (dashed line) during hydrogenation and dehydrogenation. The scattered intensities at resonance peak positions of $P_1$ and $P_2$ are used to track the dynamic processes, respectively. $P_1$ can be switched off/on through hydrogenation/dehydrogenation, whereas $P_2$ stays at the 'on' state. **(D)**

Schematic of the optical setup for capturing the holographic images. QWP, quarter waveplate; LP, linear polarizer. **(E)** Representative snapshots of the holographic images during hydrogenation and dehydrogenation. 'Harmony' (left) can be switched off/on using $H_2$/$O_2$, whereas 'peace' (right) stays still.

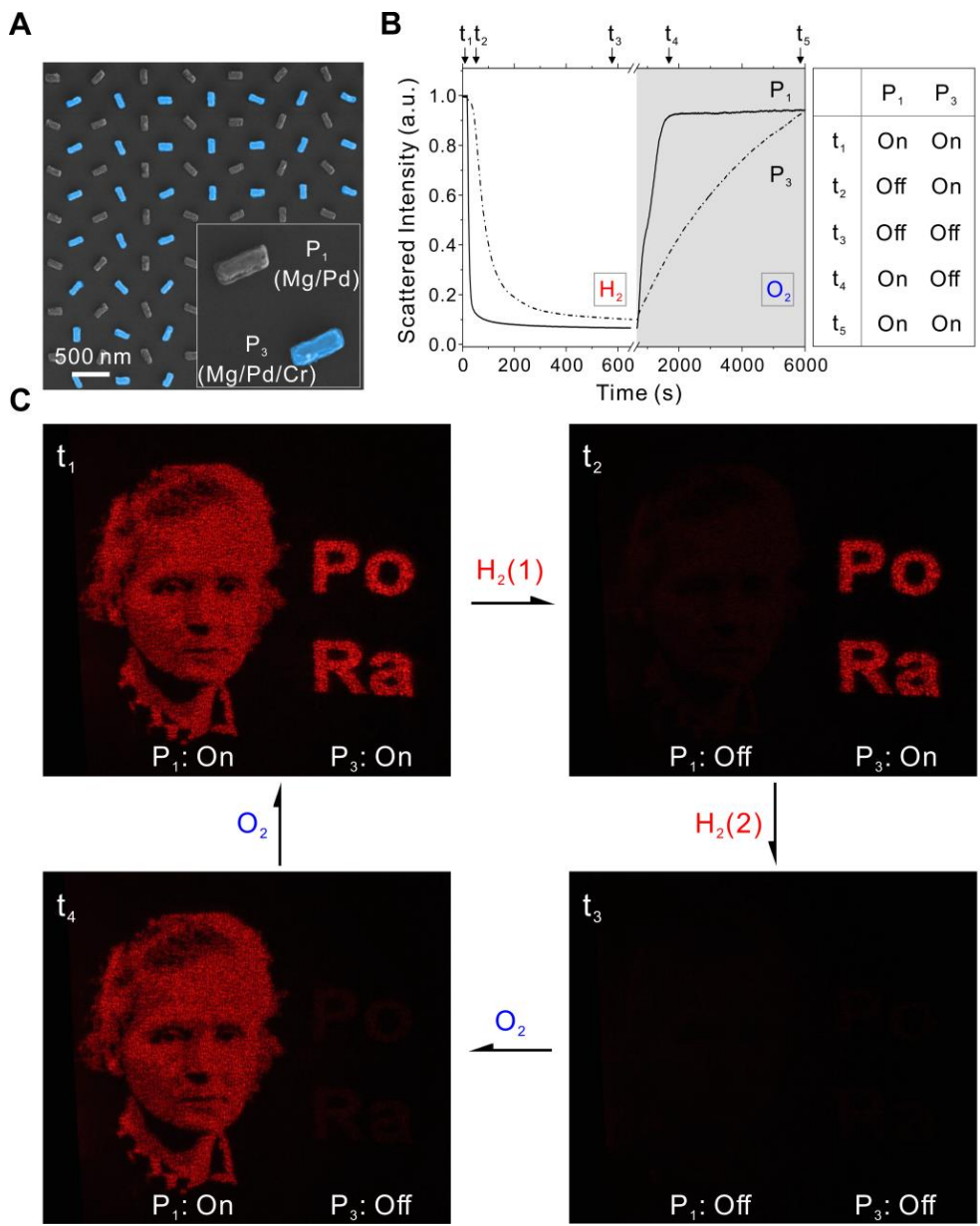

**Fig. 3. Metasurface hologram with differently sequenced dynamics.** (**A**) Mg/Pd (P$_1$) and Mg/Pd/Cr (P$_3$) nanorods work as dynamic pixels with different reaction kinetics. Overview SEM image of the hybrid plasmonic metasurface as well as the enlarged SEM image of the unit cell (600 nm × 600 nm). The Mg/Pd/Cr (P$_3$) nanorods are represented in blue color. (**B**) Time evolutions of P$_1$ (solid line) and P$_3$ (dash-dotted line) during hydrogenation and dehydrogenation. The scattered intensities at resonance peak positions of P$_1$ and P$_3$ are used to track the dynamic processes, respectively. (**C**) Representative snapshots of the holographic images

during hydrogenation and dehydrogenation. $H_2$ (1) or $H_2$ (2) refers to a hydrogenation process until the first or the second feature completes transition. Both the portrait of Marie Curie as well as the chemical elements Po and Ra can be switched on and off independently.

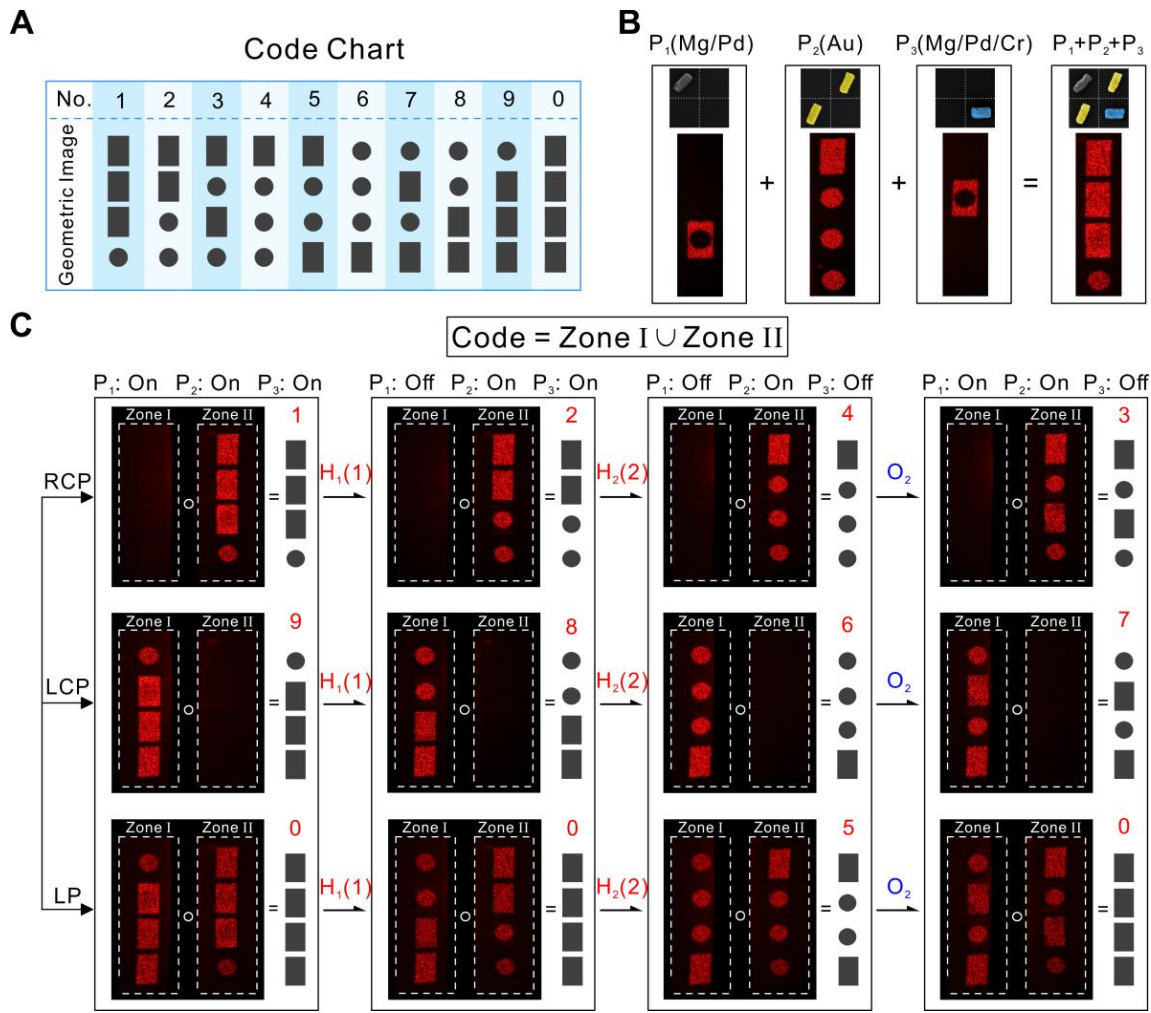

**Fig. 4. Metasurface-based dynamic geometric codes.** (**A**) Chart of the geometric codes for 0-9. (**B**) Reconstructed holographic image of '1'. Each Arabic number in the chart can be expressed by a specific holographic image reconstructed from the same metasurface multiplexed by three sets of pixels, $P_1$ (Mg/Pd), $P_2$ (Au), and $P_3$ (Mg/Pd/Cr). In order to achieve holograms with uniform intensity distributions, $P_2$ is utilized twice in each unit cell. (**C**) Helicity of light, $H_2$, $O_2$, and reaction duration constitute multiple keys to generate the numbers 0-9, which are transformable with the same metasurface. The centrosymmetry point is indicated by a white circle in each plot, highlighting the location of the zero-order reflected light. A coding rule is defined that each number corresponds to a union set of the two holographic patterns in zones I and II.

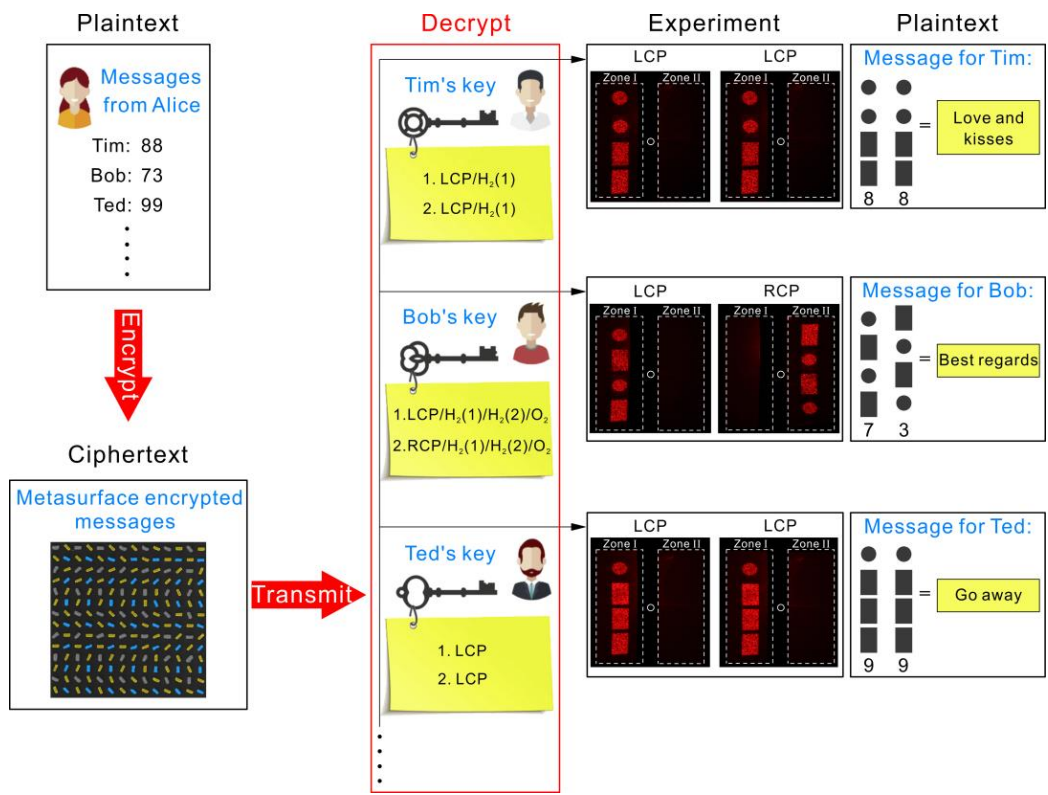

**Fig. 5**. **Proof-of-concept experiment for optical information encryption and decryption.** Alice encrypts her messages for Tim (88), Bob (73), Ted (99), and so forth all on one metasurface. Identical samples are sent to multiple receivers, who decrypt their respective messages according to the customized keys. The keys contain customer-specific information on the helicity of light (LCP, RCP, or LP), hydrogenation ($H_2$), dehydrogenation ($O_2$), and reaction duration. Different messages can be read out, corresponding to 'love and kisses' for Tim, 'best regards' for Bob, and 'go away' for Ted, and so forth in Morse codes.